# Communicability Graph and Community Structures in Complex Networks

Ernesto Estrada,[1-4*] Naomichi Hatano[4]

[1]Department of Mathematics, [2]Department of Physics, [3]Institute of Complex Systems at Strathclyde, University of Strathclyde, Glasgow G1 1XH, UK

[4]Institute of Industrial Science, University of Tokyo, Komaba, Meguro, 153-8505, Japan

---

[*] Corresponding author. Email: ernesto.estrada@strath.ac.uk

**Abstract**

We use the concept of the network communicability (Phys. Rev. E **77** (2008) 036111) to define communities in a complex network. The communities are defined as the cliques of a "communicability graph", which has the same set of nodes as the complex network and links determined by the communicability function. Then, the problem of finding the network communities is transformed to an all-clique problem of the communicability graph. We discuss the efficiency of this algorithm of community detection. In addition, we extend here the concept of the communicability to account for the strength of the interactions between the nodes by using the concept of inverse temperature of the network. Finally, we develop an algorithm to manage the different degrees of overlapping between the communities in a complex network. We then analyze the USA airport network, for which we successfully detect two big communities of the eastern airports and of the western/central airports as well as two bridging central communities. In striking contrast, a well-known algorithm groups all but two of the continental airports into one community.

# 1 Introduction

One of the most active fields of research in the study of complex networks is the detection and analysis of network communities [6, 17]. Since the seminal paper of Girvan and Newman [11], there have been many different approaches reported in the mathematical, computer sciences and physics literature dedicated to this problem [4, 12, 7, 19, 20, 25]. Communities are structural subunits in networks which are a signature of the hierarchical nature of complex systems [20]. They appear in a wide variety of systems ranging from functionally related proteins to social groups. The unambiguous identification of this *a priori* unknown structural groups in networks depends very much on how a *community* is defined.

Intuitively, a community is a group of nodes in the network which is "more densely" connected among them than with the rest of the nodes. Then, the various methods available in the literature differ mainly in the way in which they define what "more densely" connected means as well as in the algorithm that is used to find such groups of nodes. In particular, there has been a long tradition in statistical and data mining sciences in finding clusters in data, which has given rise to several clustering methods [10]. The Kernigham-Lin algorithm [14] used in computer science maximize a quality function that relates the number of edges inside each group to the number between groups. This method has inspired many other methods currently available for community detection in complex networks, which are based on optimization of certain parameters related to a group of nodes. The most popular of such methods are the ones based on modularity optimization [18]. On the other hand, the method proposed by Girvan and Newman [11] finds communities on the based on the concept of the betweenness centrality, which is one of the many centrality measures used to characterize the relevance of nodes in a complex network [26]. Another approach, which differs significantly from the previous ones, was introduced by Palla *et al*. in 2005 [20]. They use the *k*-clique percolation method to find overlapped communities in a complex network. Finally, it is necessary to mention a series of methods based on spectral techniques, which are known as spectral partitioning methods [9]. These approaches use information related to the eigenvalues and eigenvectors of matrices representing the network in order to divide it into different clusters. Other methods were reviewed in a recent literature; the reader is referred to these works and the references cited therein ref. [22].

In our recent work [8], we sketched a method of community detection in a complex network. This approach is based on the concept of the communicability between nodes in a complex network. We then defined a community as a group of nodes having larger internal communicability than the external one. The communicability is a measure of how two nodes in a network are tied to each other. It is a broad generalization of the concept of the shortest path between two nodes. When applied to the detection of communities, the communicability permits to overcome some difficulties found by the use of previously proposed methods. One of the difficulties is the use of some empirical parameters in the definition of communities. On the basis of the communicability, in contrast, the communities in a complex network can be unambiguously defined. This avoids the situation arising in many cases where the number of communities detected in a complex network depends on empirical parameters, which can yield arbitrary results. Another desirable property of algorithms of detecting communities is their ability to identify overlap between communities. As remarked by Palla *et al*. [20], we are at the same time member of several communities that overlap with each other forming a sort of supercommunities.

Here we introduce a new concept, *the communicability graph*. Communities of a graph are then re-defined as cliques of the communicability graph associated to the graph in question. In contrast to Palla *et al.*'s method [20], our method finds the cliques in this communicability graph, not the cliques existing in the original network. This method is a variation of the complete linkage method, which according to Newman [17] has perhaps more desirable properties, but it is rarely used, due to the fact that "finding cliques in a graph is a hard problem". The other difficulty stated by Newman [17] is that "the cliques are, in general, not unique". That is, a vertex can belong to two or more different cliques at the same time.

We show here that some variations of the classical algorithms of finding cliques are very efficient in very large graphs, which practically solves the first difficulty mentioned above. Then, we take advantage of the second difficulty in order to detect the overlaps among communities, which are determined as the subsets of nodes that are in more than one clique at the same time. Consequently, we find here a straightforward and unambiguous identification of the communities in a complex network. We can then easily identify all the communities in a network as well as their overlaps without any fitting or empirical parameters.

The present paper is organized as follows. In section 2, we give a short review of the concept of the communicability of a complex network. We then go to an efficient algorithm of community detection in section 3. Section 4 introduces the generalized communicability with a variable parameter, which we refer to as the inverse temperature, and thereby discusses the generalization of our algorithm of community detection using the generalized communicability. Then, we propose in Section 5 a method of analyzing communities at different degrees of overlapping. All these concepts are tested in one simple real-world network in Section 6 and finally in Section 7 we show numerical examples for the real-world complex transportation network of airports in USA.

**2 Communicability: a short review**

In order to introduce the concept of the communicability graph, we should first review the concept of the communicability itself, which we introduced in a recent work [8]. In the following, we consider simple graphs $G = (V, E)$, that is, graphs having $|V| = n$ nodes and $|E| = m$ links, without self-loops or multiple links between nodes [13]. The communicability between a pair of nodes is defined as a weighted sum of the number of all walks connecting the pair of nodes, where a walk of length $k$ is a sequence of (not necessarily different) vertices $v_0, v_1, \cdots, v_{k-1}, v_k$ such that for each $i = 1, 2 \cdots, k$ there is a link from $v_{i-1}$ to $v_i$. These walks communicating two nodes in the network can revisit nodes and links several times along the way. Let $A(G) = A$ be the adjacency matrix of the graph whose elements $A_{ij}$ are ones or zeroes if the corresponding nodes $i$ and $j$ are adjacent or not, respectively. Then, the moment $\mu_k(p,q) = (A^k)_{pq}$ gives the number of walks of length $k$ starting at the node $p$ and ending at the node $q$ [5].

As mentioned above, we define the communicability between a pair of nodes $p$ and $q$ as a weighted sum of the moments $\mu_k(p,q)$. The weight is given in a way that shorter walks receive more weights than longer ones. Specifically, we used the weight $1/k!$ for the $k$ th moment and arrived at a formula expressing the communicability $G_{p,q}$ between nodes $p$ and $q$ in terms of graph spectral parameters [8],

$$G_{p,q} = \sum_{k=0}^{\infty} \frac{\mu_k(p,q)}{k!} = \sum_{k=0}^{\infty} \frac{(A^k)_{pq}}{k!} = \sum_{j=1}^{n} \phi_j(p)\phi_j(q)e^{\lambda_j}, \tag{1}$$

where $\phi_j(p)$ is the $p$ th component of the $j$ th eigenvector of the adjacency matrix $\mathbf{A}$, which is associated with the eigenvalue $\lambda_j$. (We will generalize the weight factor $k!$ in Section 4.) We call the eigenvalues of the adjacency matrix in the non-increasing order $\lambda_1 \geq \lambda_2 \geq \cdots \geq \lambda_n$, the spectrum of the graph [5].

The communicability can be decomposed into several terms as

$$G_{p,q} = \left[\phi_1(p)\phi_1(q)e^{\lambda_1}\right]$$
$$+ \left[\sum_{2 \leq j \leq n}^{++} \phi_j(p)\phi_j(q)e^{\lambda_j} + \sum_{2 \leq j \leq n}^{--} \phi_j(p)\phi_j(q)e^{\lambda_j}\right] \quad (2)$$
$$+ \left[\sum_{2 \leq j \leq n}^{+-} \phi_j(p)\phi_j(q)e^{\lambda_j} + \sum_{2 \leq j \leq n}^{-+} \phi_j(p)\phi_j(q)e^{\lambda_j}\right],$$

where the $+/-$ signs in the sums indicate that the summation is carried out for the positive/negative components of the corresponding eigenvector, respectively. To be more precise, the summation $\Sigma^{++}$ is taken over terms in which $\phi_j(p) > 0$ and $\phi_j(q) > 0$, whereas the summation $\Sigma^{+-}$ is taken over terms in which $\phi_j(p) > 0$ and $\phi_j(q) < 0$, and so forth. Note that we separated the contribution of the principal eigenvector $\phi_1$; according to the Perron-Frobenius theorem, all components of the principal eigenvector are positive.

It may be easier to understand the communicability (2) in terms of the Green's function of a spring network. Suppose that the nodes of the network in question are balls and the links are springs. We argued [8] that the adjacency matrix $A$ is equivalent to the Hamiltonian of the spring network and the communicability (2) is equivalent to the Green's function of the spring network. The eigenvectors of the adjacency matrix represent vibrational modes of the network, and the principal eigenvector $\phi_1$, in particular, represents the translational movement of the whole network. Hence the first bracketed term on the right-hand side of the Green's function (2) represents the movement of all the nodes (the balls) in one direction after an impact on one node, as if they were part of a giant cluster formed by the whole. We are not interested in this contribution because we want to analyze the inner structure of the network. We will therefore subtract this term in the following.

In the second bracketed term on the right-hand side of Eq. (2), the nodes $p$ and $q$ have the same sign of the corresponding eigenvector (positive or negative); if we put an impact on the ball $p$, the ball $q$ oscillates in the same direction as the ball $p$. We thus regard that $p$ and $q$ are in the same cluster if there are more than one cluster in the network. Consequently, we call this second term of Eq. (2) the *intracluster communicability*. The last bracketed term of Eq. (2), on the other hand, represents an uncoordinated movement of the nodes $p$ and $q$, i.e., they have different signs of the eigenvector component; if we put an impact on the ball $p$, the ball $q$ oscillates in the opposite direction. We regard that they are in different clusters of the network. Then, we call this third term of Eq. (2) the *intercluster communicability* between a pair of nodes.

As mentioned above, we leave out the first term from Eq. (2) because we are not interested in the translational movement of the whole network and thereby consider the quantity

$$\Delta G_{p,q} = \left[ \sum_{2 \leq j \leq n}^{++} \phi_j(p)\phi_j(q)e^{\lambda_j} + \sum_{2 \leq j \leq n}^{--} \phi_j(p)\phi_j(q)e^{\lambda_j} \right]$$
$$+ \left[ \sum_{2 \leq j \leq n}^{+-} \phi_j(p)\phi_j(q)e^{\lambda_j} + \sum_{2 \leq j \leq n}^{-+} \phi_j(p)\phi_j(q)e^{\lambda_j} \right] \quad (3)$$
$$= \sum_{j=2}^{\text{intracluster}} \phi_j(p)\phi_j(q)e^{\lambda_j} - \left| \sum_{j=2}^{\text{intercluster}} \phi_j(p)\phi_j(q)e^{\lambda_j} \right|,$$

where in the last line we used the fact that the intracluster communicability is a positive term and the intercluster communicability is a negative one [8]. We thereby defined [8] that the nodes $p$ and $q$ belong to the same cluster if the quantity $\Delta G_{p,q}$ is positive, and they do not if it is. In other words, we defined unambiguously a community $C$ of a network $G = (V, E)$ as follows:

**Definition 1:** $C \subseteq V$ is a community of $G$ if, and only if, $\Delta G_{p,q}(\beta) > 0 \quad \forall (p,q) \in C$.

This definition contains the seed of an algorithm of detecting the communities in a complex network. The objective of this algorithm is to identify all pairs of nodes having $\Delta G_{p,q}(\beta) > 0$ in the graph. For the purpose, we will introduce in the next section a new concept that permits the identification of network communities in an efficient way.

## 3 Communicability graphs

In the present section, we define the communicability graph for a complex network on the basis of Eq. (3). Before introducing the definition, let us explain the motivation. The introduction of the concept of communicability leads invariably to an unambiguous definition of communities in a complex network. However, this does not mean that we have an elaborate algorithm of detecting such communities in the network. The problem consists of identifying all pairs of nodes having $\Delta G_{p,q}(\beta) > 0$ in the graph. This can be done, in principle, by carrying out certain arrangements of the matrix representing the values of $\Delta G_{p,q}(\beta)$ for all pairs of nodes in the graph. By this way the communities can be identified as the positive submatrices in such a matrix. However, this method faces the problem of identifying nodes which are in more than one community at the same time due to the community overlapping. A better solution to this problem consists of representing the values of $\Delta G_{p,q}(\beta)$ between every pair of nodes as a new graph. Then, we can find communities as cliques in this graph associated to the network in question. We call this graph, for obvious reasons, the *communicability graph* of the complex network.

Let us define the communicability graph for a complex network. First, we introduce the following function,

$$\Theta(x) = \begin{cases} 1 & \text{if } x \geq 0, \\ 0 & \text{if } x < 0. \end{cases} \quad (4)$$

Let $\Delta(G)$ be a matrix whose $(p,q)$ entries are given by $\Delta G_{p,q}(\beta)$.

**Definition 2:** The communicability graph $\Theta(G)$ of the graph $G$ is the graph whose adjacency matrix is given by $\Theta(\Delta(G))$, where $\Theta(\Delta(G))$ results from the elementwise application of the function $\Theta(x)$ to the matrix $\Delta(G)$. The nodes of $\Theta(G)$ are the same as the nodes of $G$, and two nodes $p$ and $q$ in $\Theta(G)$ are connected if, and only if, $\Delta G_{p,q}(\beta) > 0$ in $G$.

Now, let us recall the following concepts from graph theory [13]. A *complete subgraph* is a part of a graph in which all nodes are connected to each other. A *clique* is a locally maximal complete subgraph. Then, using the definitions 1 and 2, it is straightforward to realize that every community in the graph $G$ is represented by a clique of the accompanying communicability graph $\Theta(G)$. Thence the problem of finding the communities of a graph $G$ is reduced to the problem of finding the cliques of $\Theta(G)$.

The enumeration of all cliques in a graph is known as the *all-clique problem* [2, 21]. That is, given a graph we need to determine all maximal complete subgraphs. This problem is a very well-known NP-hard problem. A classic branch-and-bound approach for solving this problem is the Bron-Kerbosch algorithm (*BK-algorithm*) [3]. The BK-algorithm works recursively and is reported as the fastest enumeration algorithm. It is also robust and easily modifiable [15], which makes it a good candidate to be applied for large complex networks. The algorithm finds all cliques in a graph exactly once, using three sets $P$, $Q$, and $R$. In Table 1, we give a general algorithm for identifying the communities in a complex network based on the cliques of the communicability graph.

Table 1. Algorithm for identifying communities in a network.

---

GENERATE COMMUNICABILITY GRAPH, $\Theta(G)$

ENUMERATE CLIQUES $(P,Q,R)$

    ▷ enumerates all cliques in the communicability graph $\Theta(G)$

    $P$: set of nodes belonging to the current clique

    $Q$: set of nodes which can be added to $P$

    $R$: set of nodes which are not allowed to be added to $P$

    $N[p]$: set of neighbors of node $p$ in $G$

01.  Let $Q$ be the set $\{p_1,\cdots,p_k\}$;

02.  **if** $Q = 0$ and $R = 0$

03.    **then** REPORT-CLIQUE;

04.    **else for** $i \leftarrow 1$ to $k$

05.      **do** $Q \leftarrow Q \setminus \{p_1\}$;

06.        $Q' \leftarrow Q$;

07.        $R' \leftarrow R$;

08.        $N \leftarrow \{v \in V | \{p_i, q\} \in E\}$;

09.        ENUMERATE-CLIQUES $(C \cup \{p_i\} P' \cap N, R' \cap N)$;

10.        $R' \leftarrow R \cup \{p_i\}$;

11.      **od**;

12.  **fi**;

---

In a recent analysis of the complexity for generating all maximal cliques in a graph using pruning methods as in the BK-algorithm, Tomita et al. [24] have found that the worst-case time is $O(3^{n/3})$ for an $n$-node graph. However, in the same study these authors carried out a computational analysis which is of greater importance for our current purposes. They showed that the algorithm runs very fast in practice. For instance, for a random graph having $n = 10,000$ nodes and average degree $k = 10$, the algorithm consumes 10.86 seconds in Pentium4 2.20 GHz CPU with 2GB main memory and a Linux operating system. In fact, the graph contains 49,738 maximal cliques. Even for a random graph having $n = 10,000$ nodes and a very large average degree $k = 1000$, the algorithm consumes 1,825.45 seconds in identifying 229,786,397 cliques. It is useful to know that most of the real-world networks which have been already analyzed in the literature are sparse and have average degree $k < 500$ (see for instance ref. [16]). In Fig. 1 we plot some of the results obtained by Tomita *et al*. [24], where the CPU time obtained with the KB algorithm is plotted versus the size of sparse random graphs having from 1,000 to 10,000 nodes as well as for random graphs having 10,000 nodes and the average degree changing from 10 to almost 500.

**Insert Fig. 1 about here.**

**4 Generalized communicability and its application to community detection**

In the precedent sections we considered that the communicability in the form (1), where the weight of the summation was $1/k!$ for the moment $\mu_k$. Here we generalize the weight in order to allow a variable parameter and seek possibilities of generalized definition of the communicability.

Specifically, we use the weight $\beta^k / k!$ as

$$G_{p,q}(\beta) = \sum_{k=0}^{\infty} \frac{\beta^k}{k!} \mu_k(p,q) = \sum_{k=0}^{\infty} \frac{\beta^k \left(A^k\right)_{pq}}{k!} = \left(e^{\beta A}\right)_{pq}. \tag{5}$$

This is reduced to the original definition at $\beta = 1$. The best way of thinking about this parameter may be to suppose that we submerge the network into a thermal bath at the temperature $T$, where $\beta = (k_B T)^{-1}$ with $k_B$ a constant. The thermal bath represents an external situation which affects all the links in the network at the same time. For instance, if we think about a protein-protein interaction network the thermal bath can represent a level of stress that the cell suffers due to external conditions. Then, these external factors affect all protein-protein interactions at the same time. We thereby call the parameter $\beta$ the *inverse temperature* hereafter.

For $\beta \ll 1$, the longer walks between the two nodes are more severely penalized than in the definition (1); only very short walks are accounted as the generalized communicability between the nodes. As $\beta \to 0$, the generalized communicability is approximately proportional to the number of the shortest paths, a measure used in some works [11] to find communities in a complex network. When $\beta \gg 1$, the long walks receive more weights, indicating that we take more account of long walks in considering communities.

The inverse temperature $\beta$ can have different meanings in different contexts. It can represent different levels of stress of a cell or social agitation in a social network. Another way of seeing the generalized communicability (5) is to regard $\beta$ as the strength of each link. Consider a generalized adjacency matrix $A(\beta)$, whose element $(A(\beta))_{ij}$ is $\beta$ instead of unity if the nodes $i$ and $j$ are adjacent to each other. Then the generalized communicability (5) is simply given by

$$G_{p,q}(\beta) = \sum_{k=0}^{\infty} \frac{(A(\beta)^k)_{pq}}{k!}, \tag{6}$$

which goes back to the definition (1). In this context, by changing the inverse temperature $\beta$, we change the communication strength of the links, e.g., the spring constant of a spring network, the conductivity of a resister network, the bandwidth of a telephone network and so on. In the extreme case $\beta = 0$, the network behaves as an empty graph, i.e., a graph without links. For large $\beta$, on the other hand, communication between nodes takes place by using long-range routes.

Let us discuss Eq. (5) further from a point of view of the network spectrum. For simple graphs the communicability between a pair of nodes in a network at inverse temperature $\beta$ is given by,

$$G_{p,q}(\beta) = \sum_{j=1}^{n} \phi_j(p) \phi_j(q) e^{\beta \lambda_j}. \tag{7}$$

**Proposition 1:** As $\beta \to 0$ ($T \to \infty$), the communicability between any pair of nodes in the graph vanishes as

$$G_{pq}(\beta \to 0) = \sum_{j=1}^{n} \phi_j(p) \phi_j(q) = 0. \tag{8}$$

**Proposition 2:** As $\beta \to \infty$ ($T \to 0$), the communicability between any pair of nodes in the graph is determined by the Perron-Frobenius eigenvalue/eigenvector of the adjacency matrix,

$$G_{pq}(\beta \to \infty) = \phi_1(p) \phi_1(q) e^{\lambda_1}$$

*Proof:* Let $\mu_r(p,q)$ be the number of walks of length $r$ between nodes $p$ and $q$. It is known that for very large $r$, the number of such walks is dominated by the principal eigenvalue of the adjacency matrix of the graph,

$$\mu_r(p,q) \approx \phi_1(p) \phi_1(q) \lambda_1^r \quad \text{as } r \to \infty. \tag{9}$$

Then in the limit $\beta \to \infty$, we can see that the communicability between the nodes $p$ and $q$ is mainly determined by the Perron-Frobenius eigenvalue and eigenvector of the graph, which proves the result,

$$G_{pq}(\beta \to \infty) = \sum_{k=0}^{\infty} \frac{\mu_k(p,q)}{k!} \approx \sum_{r=0}^{\infty} \phi_1(p)\phi_1(q)\frac{\lambda_1^r}{r!} = \phi_1(p)\phi_1(q)e^{\lambda_1} . \tag{10}$$

These two analytical results have important consequences for the detection of communities at different inverse temperatures. When $\beta \to 0$, we have Eq. (8) and hence

$$\Delta G_{p,q}(\beta \to 0) = -\phi_1(p)\phi_1(q) < 0 , \tag{11}$$

which implies that the modulus of the intercluster communicability is larger than the intracluster one for every pair of nodes,

$$\sum_{j\geq 2}^{\text{intracluster}} \phi_j(p)\phi_j(q) < \left| \sum_{j\geq 2}^{\text{intercluster}} \phi_j(p)\phi_j(q) \right| \quad \forall p \in V, q \in V . \tag{12}$$

The communicability graph $\Delta(G)$ is an empty graph and hence there is no communities in the graph when $\beta \to 0$.

The situation for the very large inverse temperatures is quite straightforward. Because $G_{pq}(\beta \to \infty) = \phi_1(p)\phi_1(q)e^{\beta\lambda_1}$, we have

$$\Delta G_{pq}(\beta \to \infty) = 0 \quad \forall p \in V, q \in V , \tag{13}$$

which means that every pair of nodes in the communicability graph are connected to each other, i.e., the communicability graph is a complete graph. In other words, there is one community formed by all the nodes of the graph. The study of the values $0 \leq \beta < \infty$ is carried out in a computational way in the following section. There are several other mathematical and computational aspects of this problem which, for the sake of brevity, will be not considered here but elsewhere.

**5 Mergence of overlapping communities**

One of the characteristics of the current approach is its ability to identify the overlaps among the communities in a complex network. In the present section, we discuss how to analyze the communities once we detect them by the method proposed above.

Two communities are overlapped if they share at least one common node. We can use this information in order to analyze the degree of overlapping between two communities, which can be related to the similarity between the communities in question. Then, we propose the following index as the overlap between the communities $A$ and $B$ in a network:

$$S_{AB} = \frac{2|A \cap B|}{|A|+|B|},$$

where the numerator is the number of nodes in common in the two communities and the denominator gives the sum of the number of nodes in both communities. This index is known in the statistical literature as the Sørensen similarity index [23] and is used to compare the similarity between two samples in ecological systems in particular. The index is bounded as $0 \leq S_{AB} \leq 1$, where the lower bound is obtained when no overlap exists between the two communities and the maximum is reached when the two communities are identical.

We can calculate the similarity index $S_{AB}$ for each pair of communities found in the network and then represent all results as a matrix $\mathbf{S}$. Now, let us suppose that we are interested in identifying only those communities that have an overlap lower than a certain value $\alpha$. In other words, we will be interested only in those communities having $S_{AB} < \alpha$ in the matrix $\mathbf{S}$. Then, the communities for which $S_{AB} \geq \alpha$ need to be merged together into simpler communities. For instance, if there are three communities A, B and C in a network having overlaps $S_{AB} = 0.75$, $S_{BC} = 0.85$ and $S_{AC} = 0.55$, and if we are interested in those communities having overlaps lower than $\alpha = 0.5$, the communities A, B and C need to be merged into a single community. If their overlaps were $S_{AB} = 0.75$, $S_{BC} = 0.85$ and $S_{AC} = 0.15$ for the same value of $\alpha = 0.5$, then we need to merge the communities A and B into a community AB as well as the communities B and C into a community BC. Now, we need to analyze the overlap between the newly merged communities AB and BC. If $S_{AB,BC} \geq \alpha$, then we merge the two communities into a community ABC, but if $S_{AB,BC} < \alpha$ we do not.

The general procedure of managing overlapped communities can be described as follows:

i) Find the communities in the network following the approach described in the preceding sections;

ii) Calculate $S_{AB}$ for all pairs of communities found in the previous step and build the matrix $\mathbf{S}$;

iii) For a given value of $\alpha$, build the matrix $\mathbf{O}$, whose entries are given by
$$O_{AB} = \begin{cases} 1 & \text{if } S_{AB} \geq \alpha, \\ 0 & \text{if } S_{AB} < \alpha, \text{ or } A = B; \end{cases}$$

iv) If $\mathbf{O} = \mathbf{0}$, go to the end; else go to the step (v);

v) Enumerate the cliques in the graph whose adjacency matrix is $\mathbf{O}$. Every clique in $\mathbf{O}$ represents a group of communities with overlaps larger than or equal to $\alpha$;

vi) Build the merged communities by merging the communities represented by the nodes forming the cliques found in the step (iv) and go to the step (ii);

vii) End.

**6 An illustrative example: the Zachary karate club**

In the present section, we first demonstrate the method presented in section 3. Next, we demonstrate that the inverse temperature introduced in section 4 reveals an internal structure of communities. Finally we show how to manage the overlaps among communities. For these purposes we consider a friendship network known as the Zachary karate club, which has 34 members (nodes) with some friendship relations (links) [27]. The members of the club, after some entanglement, were eventually fractioned into two groups, one formed by the followers of the instructor and the other formed by the followers of the administrator. This network has been analyzed in practically every paper considering the problem of community identification in complex networks. In Fig. 2a we illustrate the Zachary network in which the nodes are divided into the two classes observed by Zachary on the basis of the friendship relationships among the members of the club.

As mentioned earlier, in the work of Girvan and Newman [11], an almost perfect split was obtained for the two groups with the exception of the node 3, which was classified incorrectly. In the Fig. 2b, we illustrate the communicability graph $\Theta(G)$ of the Zachary network. As can be seen $\Theta(G)$ correctly divides the network into two groups. There is very high internal communicability among the members of the respective groups but there is almost no communicability between the groups. In fact, the node 3 is correctly included in the group of the instructor (node 1).

**Insert Fig. 2 about here.**

The analysis of the cliques in the communicability graph reveals a more detailed view of the community structure of this network. Accordingly, there are five different cliques representing five overlapping communities in the network. These communities are given below, where the numbers correspond to the labels of the nodes in Fig. 2a:

$A$ : {10,15,16,19,21,23,24,26,27,28,29,30,31,32,33,34} ;

$B : \{9,10,15,16,19,21,23,24,27,28,29,30,31,32,33,34\}$;

$C : \{10,15,16,19,21,23,24,25,26,27,28,29,30,32,33,34\}$;

$D : \{1,2,3,4,5,6,7,8,11,12,13,14,17,18,20,22\}$;

$E : \{3,10\}$.

As can be seen, the first three communities, which correspond to the group of the administrator (node 34), are formed by 16 members each, and display an overlap of about 94%. The fourth community corresponds to the one of the instructor (node 1) and also has 16 members. The last community is formed by the nodes 3 and 10 only. This community displays overlaps with the communities of the administrator as well as with the one of the instructor. In fact, the node 10 appears in the communities $A$ to $D$, and the node 3 appears in the communities $D$ and $E$. In other words, these two nodes form a "bridge" of the administrator followers and the instructor followers. Our approach detects this "bridge" as a community.

Now we illustrate the method of studying the overlaps among the different communities in a complex network, which was explained in the previous section. Using the information given above about the membership of every node to the different communities in the Zackary karate club network, we build the community-overlap matrix **S** for this network, which is given below:

$$\mathbf{S} = \begin{bmatrix} 1.000 & 0.938 & 0.938 & 0.000 & 0.111 \\ & 1.000 & 0.875 & 0.000 & 0.111 \\ & & 1.000 & 0.000 & 0.111 \\ & & & 1.000 & 0.111 \\ & & & & 1.000 \end{bmatrix}.$$

For the sake of simplicity, we study the communities with overlapping lower than 10% ($\alpha = 0.10$). In this case the matrix **O** is given as follows:

$$\mathbf{O} = \begin{bmatrix} 0 & 1 & 1 & 0 & 1 \\ & 0 & 1 & 0 & 1 \\ & & 0 & 0 & 1 \\ & & & 0 & 1 \\ & & & & 0 \end{bmatrix}.$$

There are two cliques in the graph represented by this adjacency matrix, which corresponds to (A, B, C, E) and (D, E). Then we merge the four communities A, B, C and E into one community as well as the two communities D and E into another. Let these two communities denoted by $C_1$ and $C_2$, respectively. Next, the overlap between these two communities is only 0.06, which is not larger than $\alpha = 0.10$. Thus the new matrix $\mathbf{O}$ is simply a zero matrix and we stop the process as indicated in the previous section. Then $C_1 = A \cup B \cup C \cup E$ and $C_2 = D \cup E$ are the two communities existing in this network with overlaps less than 10%.

Finally, we study the effect of the temperature on the structure of communities in the Zachary social network. For $\beta = 1$, we detect five communities as explained above, and in particular, four communities with three or more nodes. By changing $\beta$, we detect more or less communities. In Fig. 3 we illustrate the number of communities of different sizes existing at different values of $\beta$. As expected from our theoretical analysis in section 4 at $\beta = 0$, there are no communities as the network is formed by isolated nodes only. However, at $\beta = 0.1$ there are 9 communities with 3 or more nodes, which increases up to 25 for $\beta = 0.3$. These values of $\beta$ represent very low strength of the interaction between the nodes. At such values of $\beta$, the communities present in the network are formed by very few nodes, e.g., by three or four nodes only. As the value of $\beta$ increases, the number of communities starts to decrease. For instance, at $\beta = 0.6$ there are 11 communities and at $\beta = 0.8$ there are only 5 communities. These communities are of larger size and they are formed by fusing together the previous small communities observed at very low values of $\beta$. At $\beta = 1.0$ the network is at normal conditions, i.e., it is the network at the real-world conditions. In this case, we observe the 4 communities (of more than 3 nodes) previously detected for this social network.

If the value of $\beta$ is increased beyond 1.0, the number of communities decreases until the whole network forms only one community. For instance, at $\beta = 1.1$ there are 3 communities and at $\beta = 1.2$ only 2 communities remain. These two communities correspond to the main two factions formed in the karate club, i.e., the administrator and the trainer ones. Then, if we increase the inverse temperature over the value $\beta = 1.0$, the communities of the network at normal conditions are fused together until they form the final "all-nodes" community (for a connected network). A similar situation is observed if we analyze the communities with more than 4 or more than 5 nodes (see Fig. 3).

**Insert Fig. 3 about here.**

We can interpret the above result in two ways. First, the "inverse temperature" $\beta$ may be detecting the "depth of community ties." As explained in section 4, when $\beta$ is large, we "overestimate" the contributions of long walks between a pair of nodes; we thereby detect wide-range communities which are loosely bound internally. As we decrease $\beta$, we focus more on short-range ties and hence "dissect" the loosely bound communities. In other words, we detect tightly bound communities when $\beta$ is small. This is why the number of communities increases for small $\beta$ in Fig. 3. The decrease around $\beta = 0$ is simply caused by the fact that we "dissect" communities down to less than 3 nodes.

In this particular example of the social network, we can perhaps provide a second interpretation of the inverse temperature as the "level of stress" at which the network is subjected. For instance, the case $\beta = 0$ can represent a high level of stress, like a large social agitation. At this temperature the network structure is destroyed and every individual behaves independently. As the value of $\beta$ increases the stress at which the network is subjected decreases and several organizations of the society start to appear. In an ideal situation of no stress, $\beta \to \infty$, there is only one community in the network. Consequently, the consideration of the parameter $\beta$ permits to analyze the characteristics of the community structure of a network under different external conditions by considering that such conditions affect homogeneously to the nodes of the network.

**7 Study of the USA airport network**

In this section we study the airport transportation network in the U.S.A. in 1997 [1]. Each node of the network corresponds to an airport in the U.S.A. Two nodes are connected if there is a flight connection between the two airports as of the year 1997. The network is formed by 332 airports and 2126 flight connections. Our algorithm of detecting communities based on the communicability graph identifies 11 communities from *A* to *K*, with different degrees of overlapping. The overlap matrix is displayed graphically in Fig. 4.

**Insert Fig. 4 about here.**

Next, we merge overlapping communities by using the procedure explained in section 5. The procedure detects the existence of 5 communities for $\alpha = 0.5$. The first of these communities is produced by merging the communities *A*, *B*, *C*, *D* and *K*. The second community is formed by merging the communities *E* and *F* and the third by merging the communities *I* and *J*. The communities *G* and *H* remain as non-overlapped ones. For $\alpha = 0.10$, there remain 4 communities as now the second community is formed by merging

*E*, *F* and *I*. The final communities having less than 10% overlapping after the merging procedure are as follows:

$C_1 : \{A,B,C,D,K\}$;
$C_2 : \{E,F,H\}$;
$C_3 : \{I,J\}$;
$C_4 : \{G\}$.

The overlaps between these communities are given as follows:

$$\mathbf{S} = \begin{bmatrix} 1.000 & 0.022 & 0.022 & 0.000 \\ & 1.000 & 0.000 & 0.014 \\ & & 1.000 & 0.014 \\ & & & 1.000 \end{bmatrix}.$$

The first community is the largest one, formed by 222 airports. The second largest one is $C_4$ with 110 nodes. Both communities have no overlap, which makes them interesting for further analysis.

Indeed, the community $C_1$ is mainly formed by airports in the west and central states of the U.S.A., including Alaskan, Hawaii and pacific ocean airports. The community $C_4$, on the other hand, is mainly formed by airports on the east coast of the U.S.A., Puerto Rico and Virgin Islands. There are few airports from the central part of the U.S.A. that are grouped in the community $C_4$ and few airports on the east coast are grouped in the community $C_1$. In Figure 5a, we illustrate the status of the communities $C_1$ and $C_4$. It is clear from this figure that we succeeded in detecting the communities of the airports according to their geographical locations with a few cases of interbreeding due to large east-west communication.

**Insert Figure 5 about here.**

The communities $C_2$ and $C_3$ are smaller than the previous ones, having only 10 and 11 airports, respectively. Both communities have between 1% and 2% of overlapping with the bigger communities. They can be considered as bridges between the bigger communities, just as we found in the Zachary karate club. For instance, the community $C_2$ is formed by airports from Wiscosin, Michigan, and Illinois as well as one from New York. The community $C_3$ is formed by airports from Tennessee, Arkansas, North Carolina, Louisiana, Mississippi, and Alabama as well as one from Florida. In Figure 5b, we illustrate the geographic positions of these airports; they are all in between the two bigger communities $C_1$ and $C_4$.

We also tried to detect communities in this network by using the Newman-Girvan algorithm. We consider the existence of 2 to 10 communities. This is one of the principal disadvantages of this algorithm; it needs the input of the number of communities to be investigated prior to the analysis. The algorithm does not detect any community overlaps. The best partition made by this method, according to the value of the modularity $Q$, was for the presence of seven communities ($Q = 0.079$). Three of these communities divide the airports of Alaska into three separate groups. Another community contain airports from Mariana Islands, Guam and American Palau. The fifth community groups most of the airports of Hawaii, one from American Samoa and Johnston Atoll. The sixth contains only two airports from Indiana and the other community groups 278 airports from the continental U.S.A. This partition is drastically different from the one reached by using the concept of communicability.

In the communicability-based partition it is clear that the airports in the west coast of U.S.A. have very good communicability among them as well as with the airports in Alaska and in the pacific. The communicability between the west coast airport and the ones in the central area of the U.S.A. is also very large. On the other hand, the communicability between the airports in the east part of the U.S.A. is very large among them and it is larger than that with the west coast and the central part of the U.S.A. All these results fit our intuition very well. In striking contrast, the Newman-Girvan algorithm groups all continental airports but 2 of the six airports of Indiana in one single community. This grouping is obviously not very informative about the organisation of the airport system in the U.S.A.

These analyses do not take into account the amount of flights between airports but only the existence of the connection between the airports. Considering this additional information can change radically this analysis.

## 6 Summary

In summary, we introduced here an algorithm of community detection in complex networks, using the concept of the communicability graph. The algorithm we developed here is based on detection of the cliques in the communicability graph. No internal or empirical parameters are needed in order to find all communities in a complex network at "normal" conditions. This method also gives the overlaps among the existing communities in the graph. We then present a procedure of merging the overlapping communities.

In addition, we introduced a parameter, the inverse temperature, which allows us to understand the characteristics of the community structure of a complex network under different external conditions. These external conditions are assumed to affect all the nodes of the network in a similar way and they can represent different conditions in different contexts. In the case of social networks, in particular, the inverse temperature appears to be a sort of the stress at which the society is subjected. The number and size of the communities change systematically with the changes of the inverse temperature of the network.

After demonstrating our algorithm of community detection in the Zachary karate club, we analyzed the airport network in the U.S.A. After proper merging, we successfully detected four communities that are geographically clustered. This was in surprising contrast to the community detected by the Newman-Girvan algorithm.

**Acknowledgements**

EE thanks Institute of Industrial Science, University of Tokyo for a fellowship as a Visiting Researcher and for warm hospitality during April-June, 2008. EE also thanks partial financial support from the New Professor's Fund given by the Principal, University of Strathclyde.

**Figure captions**

Fig. 1. Plot of the CPU time (sec) for sparse locally random graphs with the use of the algorithm according to Tomita et al [24]. (a) CPU time as a function of the number of nodes for graphs with average degree $\langle k \rangle = 10$. (b) CPU time as a function of the average degree for graphs having 10,000 nodes. The data used in the plots is taken from ref. [24].

Fig. 2. Representation of the social network of the Zachary karate club. (a) The two factions formed after the entanglement are represented as squares or circles, respectively. (b) The communicability graph for the Zackary karate club.

Fig. 3. Plot of the number of communities of three different sizes for different values of the inverse temperature, $\beta$.

Fig. 4. Graphical representation of the overlapping matrix for the 11 communities detected by the communicability-based algorithm in the USA airport transportation network of 1997.

Fig 5. (a) Geographical locations of the communities of airports identified by the communicability-based algorithm for 10% of overlapping. The light grey area represents the locations of the airports in the community $C_1$ and the dark grey area represents the ones in the community $C_4$. Light (dark) grey dots represent the airports that are located in the western/central (eastern) states but grouped in the community $C_4$ ($C_1$). (b) A geographical representation of the airports in the communities $C_2$ (light grey) and $C_3$ (dark grey).

**a**

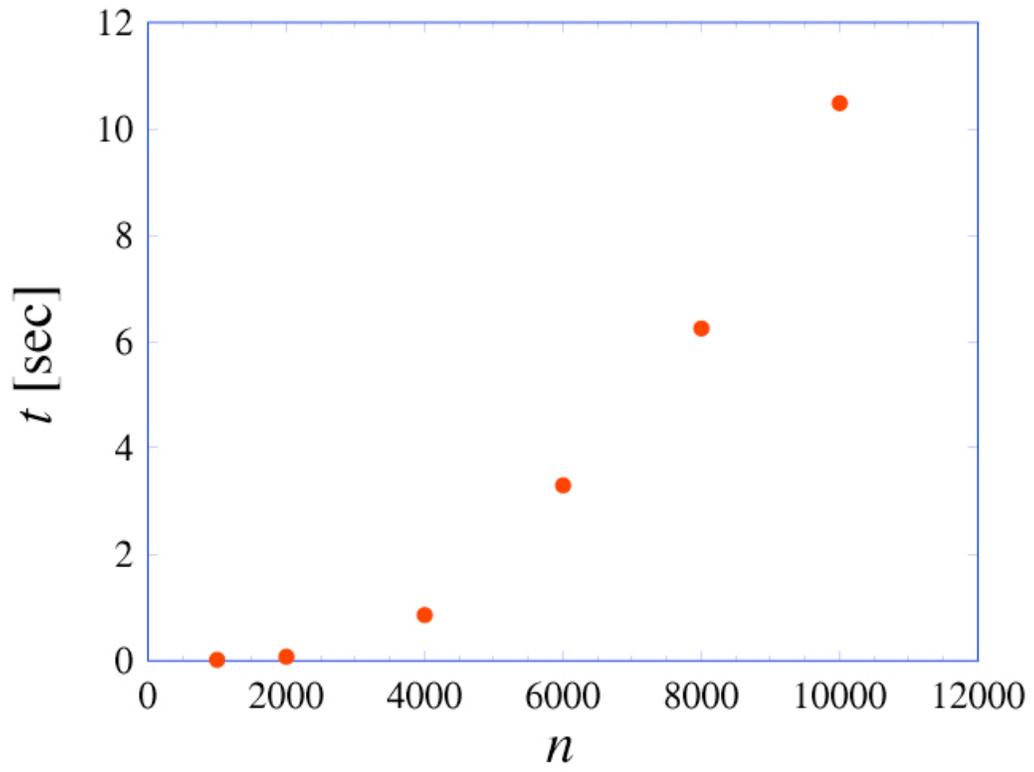

**b**

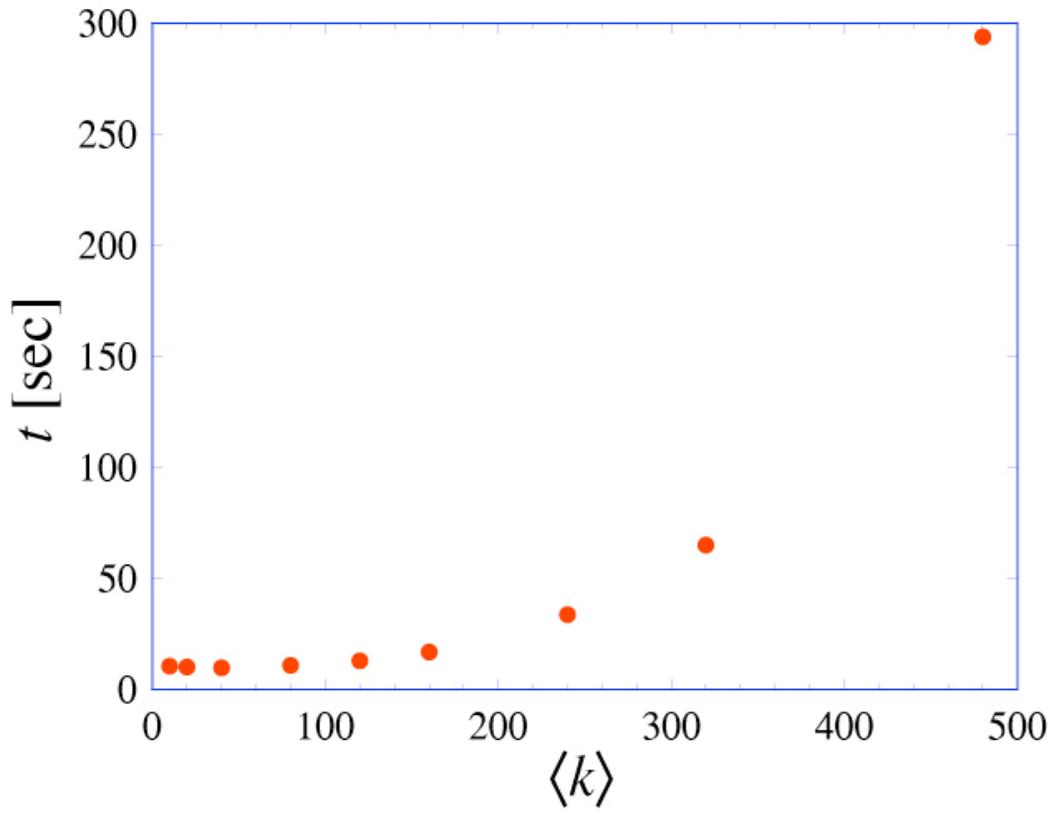

**Fig. 2**

a

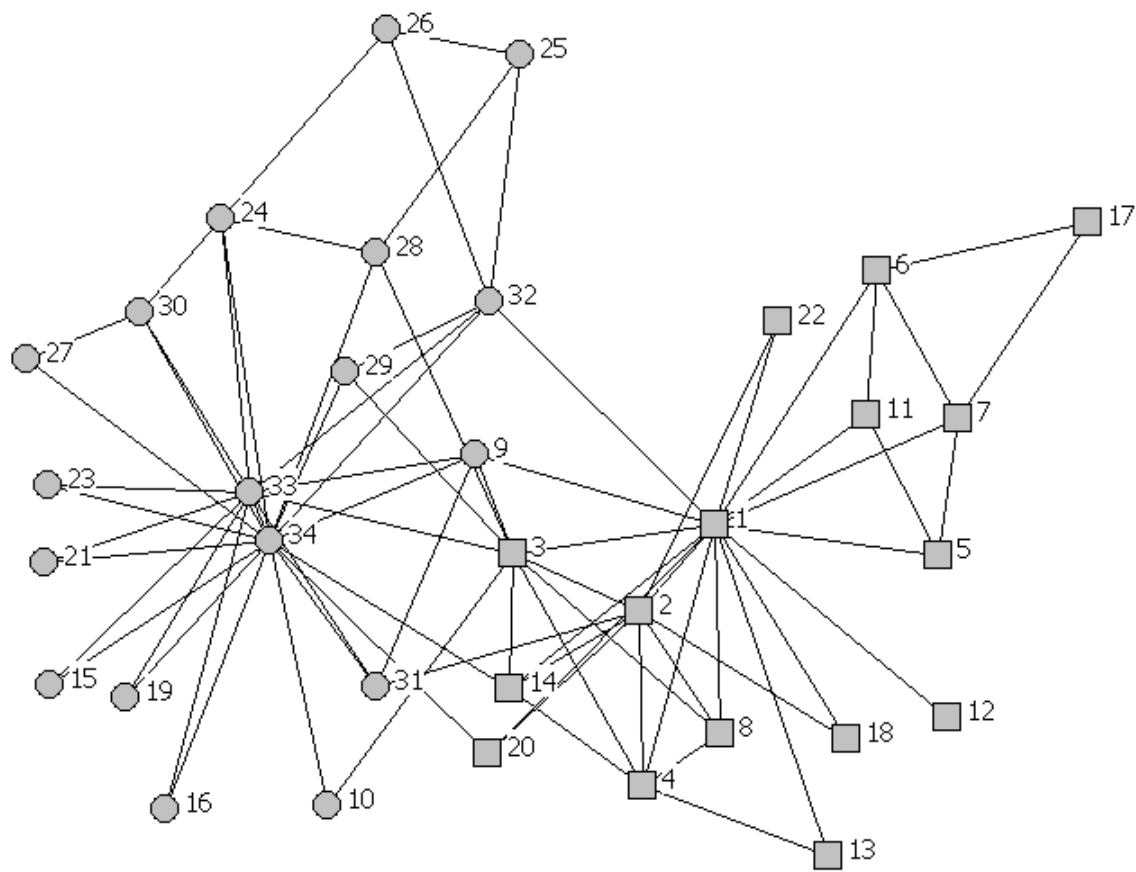

b

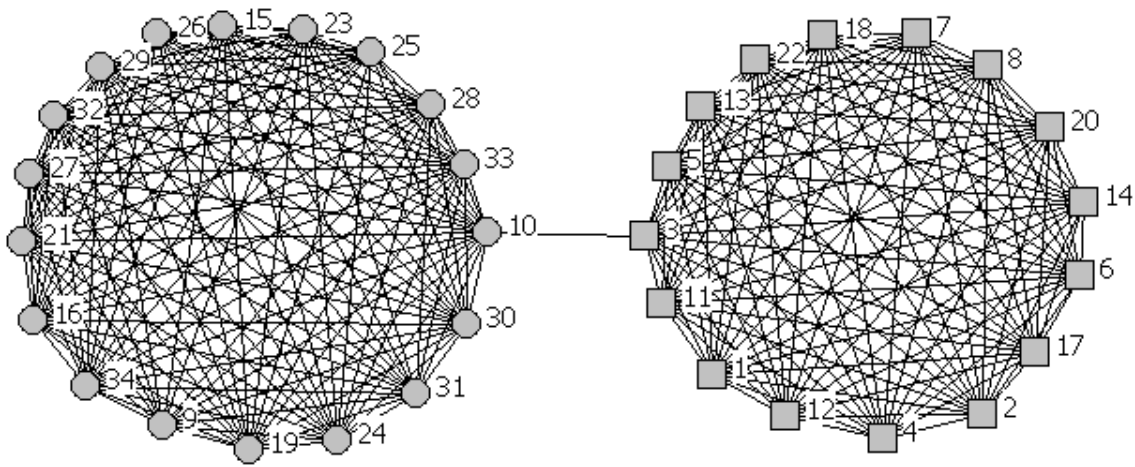

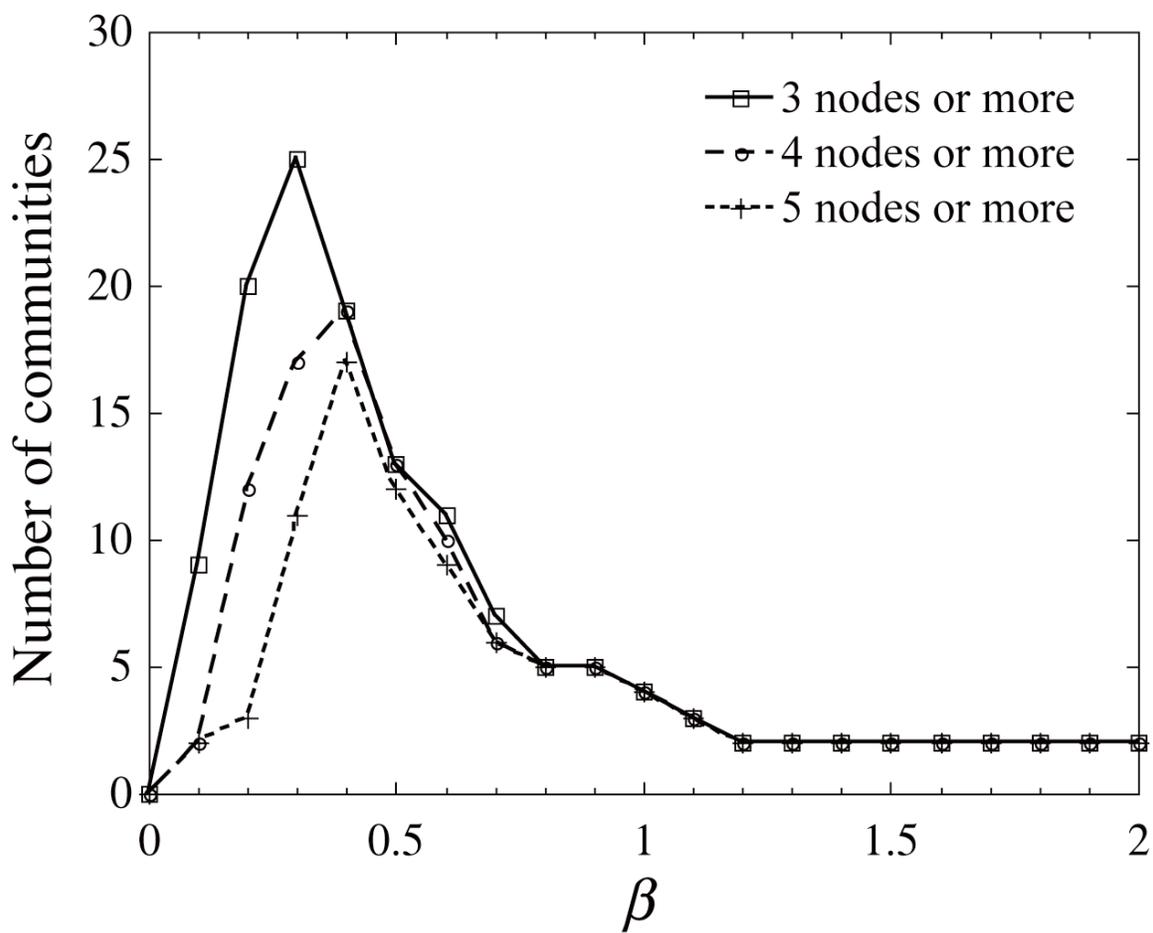

**Figure 4**

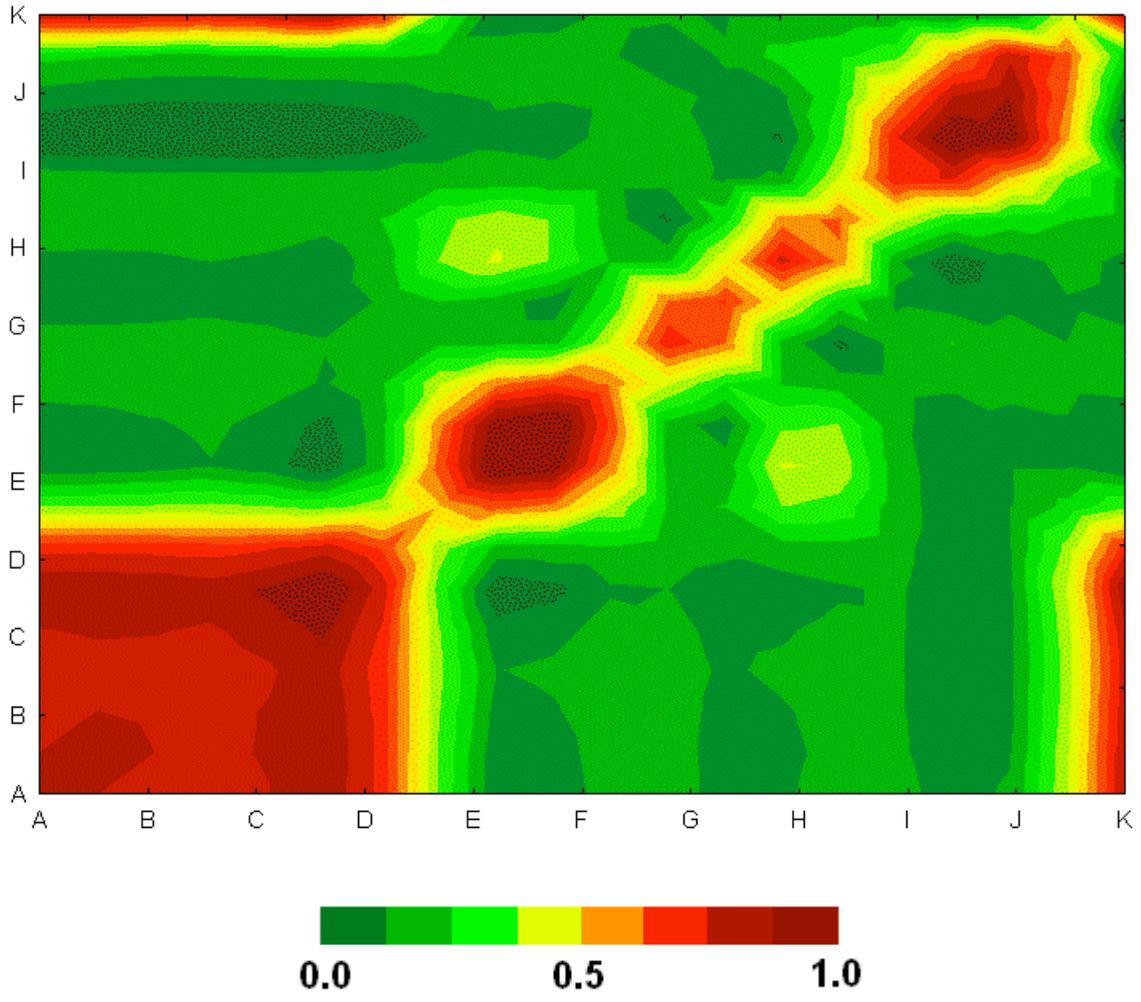

**Figure 5**

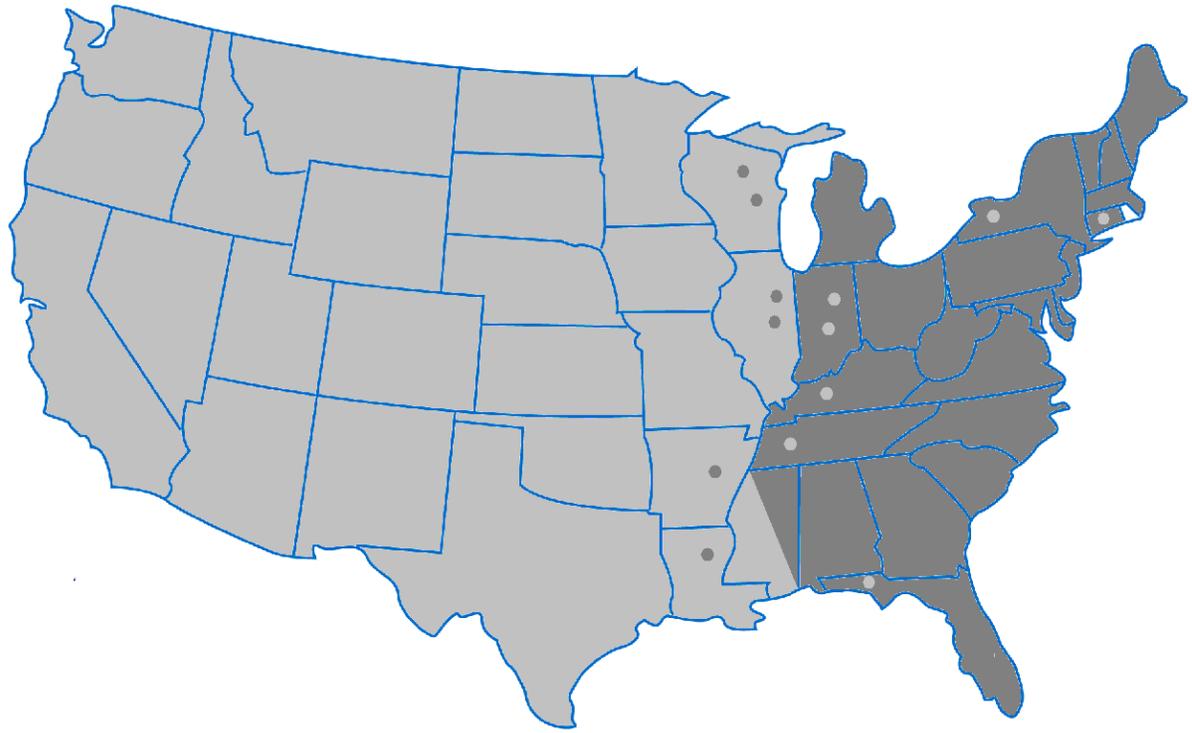

a

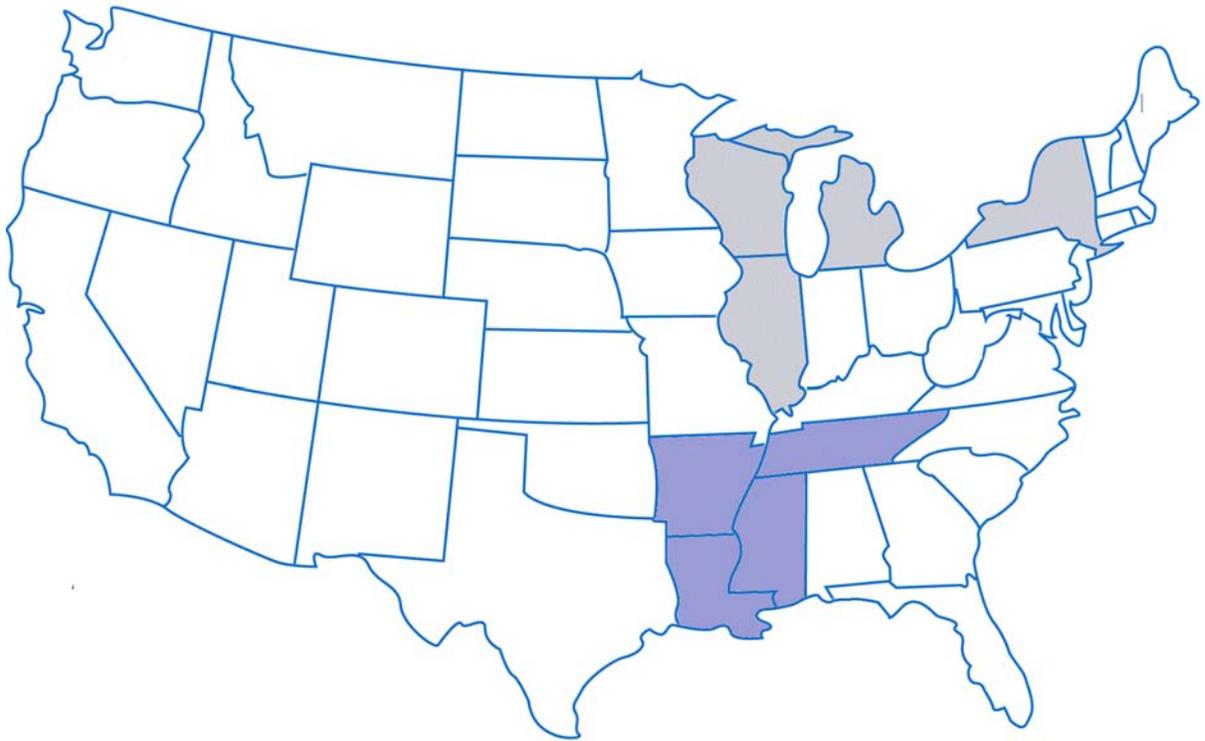

b